\documentclass[compsoc,conference,a4paper,10pt,times]{IEEEtran}
\usepackage[compress, space, sort]{cite}
\usepackage{amsmath,amssymb,amsfonts}
\usepackage{algorithmic}
\usepackage{graphicx}
\usepackage{textcomp}
\usepackage{bmpsize}
\usepackage{xcolor}

\usepackage[absolute,overlay]{textpos}

\usepackage{comment}

  \usepackage[scaled=0.8]{FiraMono} %

  \usepackage{todonotes}

  \usepackage{csquotes}

  \usepackage{listings}
  
\definecolor{key_blue}{HTML}{076d65}
\definecolor{comment_gray}{HTML}{4a4a4a}

\newcommand\YAMLcolonstyle{\ttfamily\color{black}\mdseries}
\newcommand\YAMLkeystyle{\ttfamily\color{black}\mdseries}
\newcommand\YAMLvaluestyle{\ttfamily\color{key_blue}\mdseries}

\makeatletter

\newcommand\language@yaml{yaml}

\expandafter\expandafter\expandafter\lstdefinelanguage
\expandafter{\language@yaml}
{
  keywords={true,false,null,y,n},
  keywordstyle=\color{purple},
  basicstyle=\YAMLkeystyle,                                 %
  sensitive=false,
  comment=[l]{\#},
  commentstyle=\color{comment_gray},
  stringstyle=\YAMLvaluestyle,
  moredelim=[l][\color{darkgray}]{\&},
  moredelim=**[il][\YAMLcolonstyle{:}\YAMLvaluestyle]{:},   %
  morestring=[b]',
  morestring=[b]",
  literate =    {---}{{\ProcessThreeDashes}}3
                {>}{{\textcolor{darkgray}\textgreater}}1     
                {|}{{\textcolor{darkgray}\textbar}}1 
                {\ -\ }{{\mdseries\ -\ }}3,
}

\lst@AddToHook{EveryLine}{\ifx\lst@language\language@yaml\YAMLkeystyle\fi}
\makeatother

\newcommand\ProcessThreeDashes{\llap{\color{cyan}\mdseries-{-}-}}

  \usepackage{graphicx}
  \usepackage{comment}
  \usepackage[official]{eurosym}

  \usepackage{booktabs}
  \usepackage{tabularx}
  \usepackage{listings, multicol}

  \usepackage{tikz}
  \newcommand*\emptycirc[1][0.7ex]{\tikz\draw (0,0) circle (#1);} 
  \newcommand*\halfcirc[1][0.7ex]{%
    \begin{tikzpicture}
    \draw[fill] (0,0)-- (90:#1) arc (90:270:#1) -- cycle ;
    \draw (0,0) circle (#1);
    \end{tikzpicture}}
  \newcommand*\fullcirc[1][0.8ex]{\tikz\fill (0,0) circle (#1);} 

  \newcommand*\circled[1]{\tikz[baseline=(char.base)]{
              \node[shape=circle,draw,inner sep=1pt] (char) {#1};}}

  \usepackage{pifont}

\usepackage[hyphens]{url}

\usepackage[colorlinks=true,urlcolor=black]{hyperref}
\def\BibTeX{{\rm B\kern-.05em{\sc i\kern-.025em b}\kern-.08em
    T\kern-.1667em\lower.7ex\hbox{E}\kern-.125emX}}

\usepackage{cleveref}

\begin{document}

\title{TIRA: An OpenAPI Extension and Toolbox for\\GDPR Transparency in RESTful Architectures%
}%

\author{\IEEEauthorblockN{Elias Grünewald, Paul Wille, Frank Pallas, Maria C. Borges, Max-R. Ulbricht}
\IEEEauthorblockA{\textit{Information Systems Engineering} \\
\textit{Technische Universität Berlin}\\
Berlin, Germany \\
\{eg, pw, fp, mb, mu\}@ise.tu-berlin.de} 
}

\maketitle

\begin{textblock*}{12cm}(5cm,1.5cm) %
\begin{center}
    \textit{preprint version (2021-06-10), \\accepted at 2021 International Workshop on Privacy Engineering (IWPE'21)}
\end{center}
\end{textblock*}

\begin{abstract}
  Transparency -- the provision of information about what personal data is collected for which purposes, how long it is stored, or to which parties it is transferred -- is one of the core privacy principles underlying regulations such as the GDPR. Technical approaches for implementing transparency in practice are, however, only rarely considered. In this paper, we present a novel approach for doing so in current, RESTful application architectures and in line with prevailing agile and DevOps-driven practices. For this purpose, we introduce 1) a transparency-focused extension of OpenAPI specifications that allows individual service descriptions to be enriched with transparency-related annotations in a bottom-up fashion and 2) a set of higher-order tools for aggregating respective information across multiple, interdependent services and for coherently integrating our approach into automated CI/CD-pipelines. Together, these building blocks %
  pave the way for providing transparency information that is more specific and at the same time better reflects the actual implementation givens within complex service architectures than current, overly broad privacy statements.

\end{abstract}


\section{Introduction}

Privacy regulations such as the GDPR %
prescribe several privacy principles to be complied with. One of these is the principle of transparency, which obligates controllers to inform data subjects about what data is collected, for which purposes, etc. 
Only with the appropriate information can users make well-informed and self-sovereign decisions about what information to reveal to which services.

Today, such transparency information is mostly provided in lengthy privacy policies which are typically written in legalese, hard to understand language and therefore barely read or understood by users \cite{rudolph2018}. %
In addition, information about the processing of data is kept purposely vague, to allow for flexibility regarding the implementation \cite{ThePrivacyPolicyLandscapeAftertheGDPR}. 
In their current state, most privacy policies do thus not fulfill the goal of enabling well-informed decisions and their vagueness is increasingly being challenged by authorities\footnote{See, e.g., \url{https://www.cnil.fr/en/cnils-restricted-committee-imposes-financial-penalty-50-million-euros-against-google-llc}}. As possible countermeasures, different technical approaches for providing transparency -- so called \textit{Transparency Enhancing Technologies (TETs)} -- have been proposed.
These comprise a broad variety of mechanisms, from machine-readable representations \cite{gruenewald2021, Gerl2018} to function-rich user-facing dashboards \cite{bier, Raschke2018}.

However, none of these mechanisms addresses the challenge of %
gathering the underlying transparency information within an organization in a manner that is systematic and specific enough to fulfill legal requirements. This challenge is especially relevant with the adoption of agile processes. %
Here, updates occur at an increasing frequency, are carried out by independent teams for a multitude of decoupled microservices, and are subject to a comparably low level of centralized coordination, leading to an overall system architecture that is in constant flux.

Regarding the fulfillment of legal transparency obligations, this raises the question as to how the required %
information can be collected across relevant services in a way that reflects the actual implementation of personal data processing and how this information can be integrated into a single, %
always up-to-date view that is subsequently provided to data subjects.
Established, rather manual and centrally coordinated ``inventory'' practices paradigmatically conflict with the %
goals behind agile DevOps here. %

Data controllers therefore currently face a dual challenge with regard to the provision of adequate privacy-related transparency information: First, they must be able to gather and inventory transparency-related information about the processing of personal data within their systems in line with the requirements from privacy regulations like the GDPR. Second, respective approaches must also integrate well with established principles and practices of agile, DevOps-focused engineering of complex information systems comprising a multitude of decoupled, yet interdependent microservices.%

To counter this dual challenge in real-world information systems engineering, we herein present TIRA, an OpenAPI extension and toolbox to technically address legal transparency requirements %
that consciously incorporates the givens of agile DevOps practices and RESTful architectures. In particular, we provide: %

\begin{itemize}
    \item An \textbf{OpenAPI 3 extension}, comprising a GDPR-aligned transparency vocabulary and %
    a schema extension specifying how to express %
    transparency information on a per-service basis, and %
    \item A a \textbf{set of higher-order tools} for collecting and aggregating respective information across multiple, interdependent services and for %
    integrating our approach into automated CI/CD-pipelines.%
\end{itemize}

With these contributions, we aim to %
help \textit{service providers} in capturing the processing and flow of personal data within their distributed, ever-changing systems, which is a necessary precondition for employing novel approaches that help overcome above-mentioned shortcomings of current privacy policies. Our motivation is thus to build a TET that acts from the \textit{perspective of service providers} and their developers and is actually usable to \textit{those}. This shall be done by integrating into their development routines and consciously adapting to the tools and methods employed in modern information systems engineering, imposing as little additional effort as possible. In particular, our approach closely integrates into the de-facto standard of RESTful service engineering and allows developers to easily attribute single services of a larger RESTful architecture with transparency information. On this basis, we envision the semi-automatic generation of transparency information spanning multiple services, business domains, or even companies involved in the processing of personal data %
without heavily impairing agile development routines and practices.   %

The remainder of the paper is structured as follows: we present
relevant background and related work (sec.~\ref{sec:bg_motiv}), distill requirements and delineate our general approach (sec.
\ref{sec:approach}) first. With this foundation, we introduce our transparency-related OpenAPI extension and vocabulary (sec.~\ref{sec:vocab_apispec}) and present our higher-order tools %
(sec.~\ref{sec:thub}). %
Sec.~\ref{sec:conclusion} briefly discusses our approach and concludes. 

\section{Background \& Related Work}\label{sec:bg_motiv}

Since our work is mainly motivated by the interplay between legal requirements, technical measures to fulfill them
and the actual givens of current software architectures and development practices,
these areas shall thus be briefly elaborated on.%

\subsection{Privacy, Transparency \& Technical Measures}%

\label{sec:background:sub:dp_principle}
Born as \emph{Openness Principle} in the OECD Guidelines on the Protection of Privacy in 1980~\cite{_oecd_1980}, %
transparency finds its latest incarnation as one of the core principles to which data controllers are obligated under the %
GDPR~\cite{cit_gdpr}. %
The concept of transparency in the GDPR %
aims at empowering data subjects %
regarding information asymmetries towards parties processing personal data~\cite{cit:wp29_transparency}.

Consequently, the GDPR requires numerous %
information to be provided to data subjects. %
The respective categories are distributed across Art. 13--15 GDPR and will be examined in more detail in~sec.~\ref{sec:legal_req}.
Following the concept of \emph{Data protection by design and by default}~\cite[Art. 25]{cit_gdpr}, the GDPR also obligates data controllers to establish appropriate technical measures to implement privacy principles like transparency, which 
implies the need for \emph{Transparency Enhancing Technologies (TETs)}.

Early definitions of TETs characterized them based on their means to 1) provide information about data collection, 2) act in the data subject's name to access personal data or 3) provide capabilities to lower the risk of 
profiling activities~\cite{hedbom2008survey}. %
Subsequent classification frameworks distinguish TETs %
in matters of their span width or quality of parameters such as data types presented, delivery mode, scope, or application time~\cite{cit:tet_cat_zimmermann}. %
However, there still 
exists 
a gap between legal transparency obligations 
and 
available technical measures to satisfy them,
especially concerning 
modern, real-world information systems ~\cite{spagnuelo2019accomplishing,riva2020}.%

With 
regard to microservice architectures, related work includes formalizing privacy-preserving constraints in the domain of health care \cite{formalising,theorem}. In a similar vein, specialized APIs for data transparency were proposed, that reveal \enquote{one's own user data} held by services in use \cite{theprivacyapi}. 
Besides these, however, TETs explicitly addressing RESTful microservice architectures and the interplay of multi-service environments are %
underrepresented in scientific literature.

Referring to Privacy Enhancing Technologies in general, 
the importance to build tools that include an ecosystem for their effective application and incentivize data controllers to actually use them in practice is emphasized~\cite{hansen2016data}.
This does, of course, also apply to the particular category of TETs. 
Therefore, 
solutions 
that smoothly integrate and 
align with established practices of modern information systems engineering and developers' daily routines are needed. %

\subsection{APIs, DevOps \& RESTful Architectures}
\label{sec:agile}

Currently, 
businesses increasingly adopt agile development practices, which leverage short development cycles and significantly relax on %
regimented practices, documentation, and detailed planning in favor of \textit{agile processes}.
These emphasize the importance of iterative improvement and promote greater autonomy and ownership to smaller, %
cross-functional teams \cite{estler_AgileVsStructured_2014}.

Agile development typically goes hand in hand with DevOps practices, a set of technical and cultural measures that shorten the time between committing a change and moving it to production \cite{bass_DevOpsBook_2015}. 
This reduction is enabled by continuous integration and delivery (CI/CD) pipelines that automate many software development steps, from build and test automation to automated deployment. %

Together, these practices have created an environment where updates occur at an increasing frequency and where some
of these updates may
also 
modify existing or introduce new activities of collection, processing, or even transfer of personal data.
Even if privacy requirements have so far played a minor role in the context of agile development %
\cite{agile_turn, secdevops},
this %
introduces the need to dynamically align 
provided transparency information to the continuously changing \textit{actual} %
state of the overall system.

Technology-wise, agile and DevOps-oriented practices are typically complemented by RESTful architectures comprising
dozens or even hundreds 
of microservices \cite{zimmermann_microservicesREST_2016}. %
These microservices -- %
developed and operated by the rather independent teams mentioned above -- typically expose %
APIs for representational state transfer via HTTP (so-called REST APIs). Such standardized APIs allow for 
independent development of services and the loose coupling between them required in agile practices.

To facilitate the understanding and usability of services' APIs, especially across different teams, REST APIs are typically described and documented using OpenAPI specifications\footnote{See \url{https://github.com/OAI/OpenAPI-Specification}} %
that allow each path of a REST service to be annotated with the available methods and required parameters, as well as data in- and output schemas.
This is 
usually done by service developers
in a standardized, well-structured manner directly within the program code or in accompanying specification files. %
Besides the functionality covered by the standard specification, OpenAPI can also be used for extra functionality through 
an extension mechanism\footnote{See \url{https://swagger.io/docs/specification/openapi-extensions}} 
that we will employ
for specifying transparency information within a particular service.%

\section{Requirements \& General Approach}\label{sec:approach}

From the above givens and %
their interplay in current, real world information systems engineering, we can distill a couple of requirements which shall be laid out in brief before delineating our general approach to address them. %

\subsection{Legal Transparency Obligations}\label{sec:legal_req}

As sketched in~sec.~\ref{sec:background:sub:dp_principle}, the GDPR codifies numerous transparency obligations throughout Art. 13--15. In addition, overlapping requirements concerning the maintenance of a \emph{record of processing activities} are introduced in Art. 30.
These articles therefore define which information must be provided to the data subject.

Since we aim at collecting transparency information inside distributed systems comprised of numerous microservices, we differentiate between two \emph{categories} of transparency information here. %
First, we relate comparably static information regarding the whole system as well as the company-wide implementation of GDPR requirements -- such as contact information %
of a data protection officer -- to the \emph{system (SYS)} category. %
Complementarily, 
we introduce the \emph{service (SVC)} category, covering %
all information regarding the collection of data or the processing thereof by a particular service's implementation -- such as the disclosure of personal data to specific recipients. %
Only this information is subject to the continuous change introduced by agile DevOps practices and needs to be determined dynamically by the technical mechanism to be developed while system-wide transparency information can be maintained separately (see sec.~\ref{sec:thub}).

An in-depth analysis of GDPR articles regarding transparency and said records-keeping obligations reveals
recurring patterns of information requirements. %
For instance, all examined articles (Art. 13-15 and 30) require the specification of processing purpose(s), a declaration regarding the recipients of personal data, or storage periods. %
Other aspects such as covered data categories or the origins of held data only apply in certain settings. All of these are, however, %
indispensable for legally sufficient provision of transparency information and must, therefore, be captured on a per-service level. Based on our analysis, a summary of required information %
is provided in table~\ref{tab:information-categories}.

\begin{table}[!ht]
    \caption{Categorization of transparency information required to be provided according to the GDPR.}
    \label{tab:information-categories}

\resizebox{\columnwidth}{!}{%
\centering
\tabcolsep2pt
    \begin{tabular}[]{p{0.2\textwidth}p{0.3\textwidth}}
         \toprule
         \textbf{GDPR References} & \textbf{Summary} \\
         \midrule
         \\
         \multicolumn{2}{l}{\textit{System-wide information}}\\
         \midrule
         13(1a), 14(1a), 30(1a) & Controller Contact Information\\
         13(1b), 14(1b), 30(1a) & Data Protection Officer Contact Information\\
         13(1f), 14(1f), 15(2), 30(1e) & Safeguards for third country transfer (\fullcirc)\\
         13(1c), 14(1c) & Legal basis\\
         13(1d), 14(2b) & Legitimate interest (\fullcirc)\\
         13(2b), 14(2c), 15(1e) & Right to Rectification, Deletion, \newline and Portability (\emptycirc)\\
         13(2c), 14(2d) & Right to consent withdrawal (\emptycirc, \fullcirc)\\
         13(2d), 14(2e), 15(1f) & Right to lodge complaint (\emptycirc)\\
         13(2e) & Provision mandatory (\halfcirc), \newline consequences of non-provision\\
         30(1c) & Concerned categories of data subjects\\\\
         \multicolumn{2}{l}{\textit{Service-level information}}\\
         \midrule
         13(1e), 14(1e), 15(1c), 30(1d) & Recipients\\
         13(1f), 14(1f), 15(1c), 30(1e) & Third Country / International Transfer (\halfcirc)\\
         13(1c), 14(1c), 15(1a), 30(1b) & Purpose\\
         14(1d), 15(1b), 30(1c) & Concerned categories of data\\
         13(2a), 14(2a), 15(1d), 30(1f) & Period of storage or \newline criteria to determine that period (Retention)\\
         14(2f), 15(1g) & Source / Origin of data\\
         13(2f), 14(2g), 15(1h) & Automated Decision Making / Profiling (\halfcirc), explanation\\
         \bottomrule
    \end{tabular}
}%
\newline
 $~$\\ %
\centering\textsf{\tiny{Legend: \emptycirc{} indication only, \fullcirc{} where applicable, \halfcirc{} yes/no}}

\end{table}

\subsection{Technical Requirements}\label{sec:tech_req}

While privacy (and, thus, transparency) is typically considered a non-functional requirement from the perspective of modern information systems engineering, we herein formulate our requirements from the perspective of the transparency mechanism to be developed. Like for other endeavors of privacy engineering aimed at practically applicable, and re-usable technical mechanisms (e.g. \cite{pallas-ea-2020-al-pbac,gruenewald2021}), requirements regarding the actual provision of transparency information in RESTful architectures are thus referred to as ``functional'' (FR) ones herein while those referring to other factors that influence, for instance, practical applicability are categorized as ``non-functional'' (NFR). %

On the functional side, the technical mechanism to be proposed must, first and foremost, be capable of expressing all legally required information (Req. 1, FR). The expressiveness of our mechanism is therefore explicitly aligned to the requirements from the GDPR identified above. %
In addition, we explicitly aim for a mechanism that smoothly integrates into above-mentioned, agile DevOps practices. In line with these, it is the developer of a given service who knows best what personal data the service collects, for which purposes, etc. Consequently, we strive for a service-focused approach (Req. 2, FR) that allows to harness ''service-local knowledge'' from developers in a bottom-up fashion and to integrate these into a comprehensive representation later on. For the latter, in turn, it must be possible to dynamically integrate and aggegate respective transparency information across hierarchies of decoupled and complexly intertwined services into an overall view (Req. 3, FR). %
In the light of the dynamically changing service implementations and interdependencies that characterize agile DevOps practices, this needs to be possible in an automated fashion.

Non-functional requirements, in turn, primarily regard the well-known challenge of ensuring practical applicability and fieldability of PETs ~\cite{doTheyMatch, kaminski2020recent}. In line with other work directed at comparable goals (esp. \cite{pallas-ea-2020-al-pbac, gruenewald2021}), we can at least identify the following ones: To foster practical adoption, %
the %
mechanism to be developed must integrate with well-established development practices and toolchains as coherently %
as possible (Req. 4, NFR). For similar reasons, the mechanism must be developer-friendly and raise low implementation and integration overheads (Req. 5, NFR). While developers will avoid %
hard-to-integrate technical mechanisms, %
easily usable ones will face lower refusal and may even be used out of developers' intrinsic motivation or mere curiosity \cite{pallas-ea-2020-al-pbac}.\footnote{In addition, lowering implementation overhead also works towards making the adoption of a technical mechanism legally mandatory under the GDPRs ``Privacy / Data Protection by Design'' concept \cite{edpbPbD}.} %
Finally, the required functionality should be provided in the form of re-usable artifacts (Req. 6, NFR) that encapsulate required functionality as far as possible, foster broad and consistent adoption, and ease the integration into real-world information systems \cite{pallas-ea-2020-al-pbac, gruenewald2021}. %

\subsection{Approach \& Sample Application}\label{sec:approach_approach}\label{ssec:running-example}

For fulfilling these requirements, we propose a novel approach that hooks into one of the core mechanisms and associated practices of agile, service-oriented information systems engineering: the structured description of RESTful service interfaces according to the OpenAPI standard. Such service descriptions cover a service's interface as well as it's behavior at a level of detail that allows different teams to use each other's services merely by \enquote{programming to an interface}  \cite{gamma1994design}. Our approach builds upon these OpenAPI-based capabilities for self-description and extends them to transparency-related properties of single services. This allows to grasp superior knowledge from a service's developers in a decentralized fashion (see Req. 2). Developers, in turn, do not have to adapt to paradigmatically new tools or techniques but rather stay within their well-known practice of OpenAPI-based service descriptions (see Reqs. 4 and 5). 

\begin{figure}[!t]
	\begin{center}
		\includegraphics[width=0.45\textwidth]{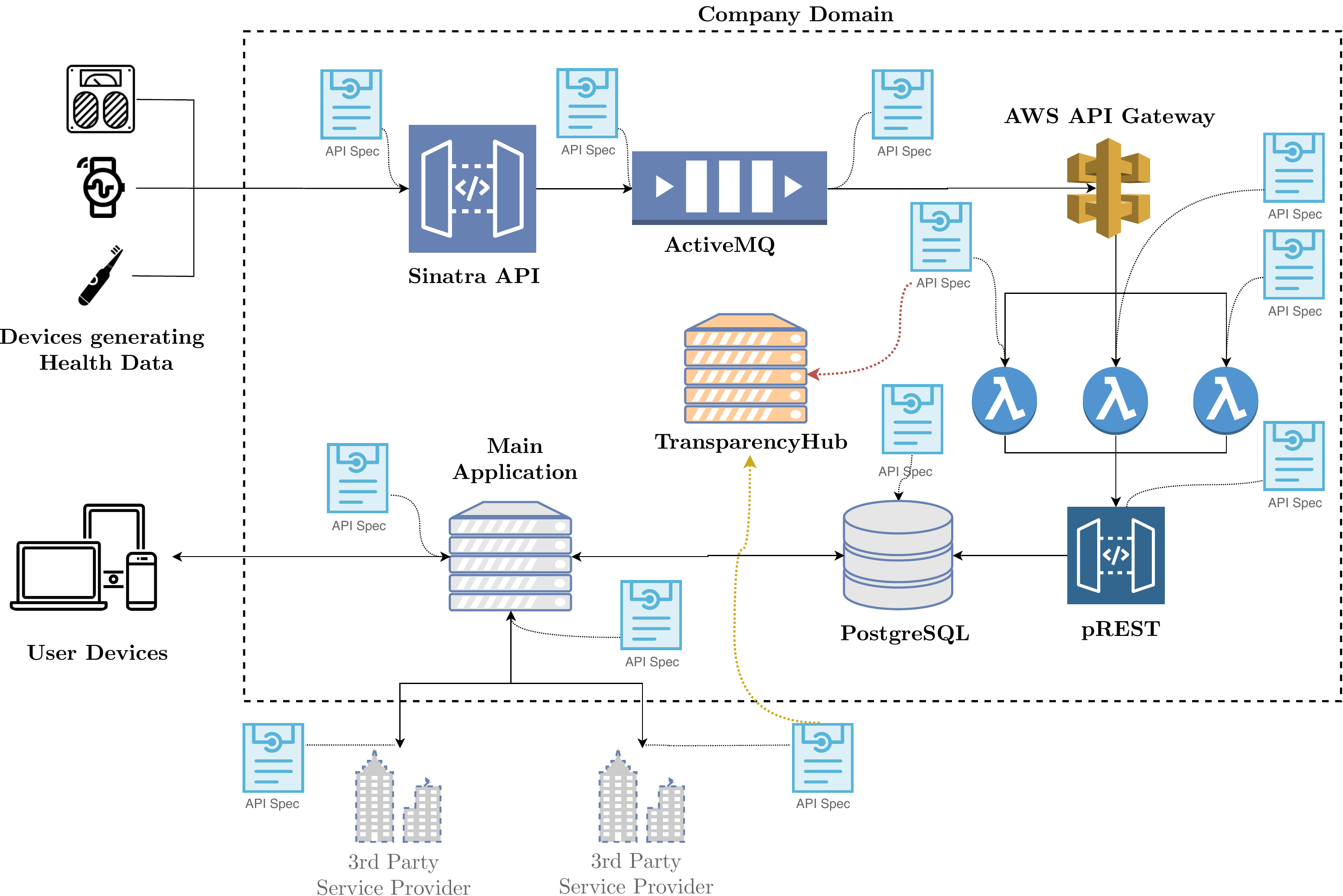}
		\caption[Overview of architectural components]
		{
		    Architectural overview of components, for which each interface is equipped with an OpenAPI specification. The red arrow, as an example, indicates the CI pipeline, which registers each version of the spec within the \textit{TransparencyHub}
		    (see sec.~\ref{sec:thub}). Third parties' service specifications can be registered manually (yellow arrow).
		}
		\label{fig:example-architecture}
	\end{center}
	\vspace{-0.5cm}
\end{figure}

Utilizing OpenAPI descriptions for the purpose of transparency does, however, require to introduce a respective OpenAPI extension and accompanying specifications particularly tailored to the GDPR's requirements (Req. 1). This %
will be done in section~\ref{sec:vocab_apispec}. Operationalizing respective descriptions in practice, in turn, requires further technical mechanisms for consolidating them into an aggregated and always up-to-date view comprising multiple services (Req. 3) and for coherently integrating respective tasks (such as notifications in case of major changes) into broadly employed CI/CD pipelines (see again Req. 4). Respective tools building upon our fundamental OpenAPI extension are briefly presented in section~\ref{sec:thub}. %

For illustrating our considerations, we assume %
a fitness application that is developed in an agile fashion, employing a multitude of paradigmatically different state of the art technologies tied together through REST interfaces. In particular, we assume activity recordings (which are to be considered personal data) to be sent from fitness-tracking devices to a RESTful API, forwarded to a publish-subscribe broker, and retrieved from there by serverless functions for validation and sanitization. Afterwards, they are inserted into a PostgreSQL database, from where they are accessed and processed by a \enquote{main application} providing a user-facing web interface etc. In addition, data might be sent to external parties (like social networks) through respective interfaces. Besides serving illustrative purposes,
we also implemented this architecture based on well-established components, allowing us to successfully validate our technical contributions (see figure~\ref{fig:example-architecture}).%

\section{A Transparency Extension and Vocabulary for OpenAPI}\label{sec:vocab_apispec}

As outlined in sec.~\ref{sec:approach_approach} above, we propose a bottom-up approach to the generation of system-wide transparency information that harnesses developers' superior knowledge about their individual services. In particular, we intend to do so through developer-provided transparency information within the respective services' OpenAPI specifications. For this purpose, we introduce a dedicated extension for OpenAPI as well as an accompanying vocabulary below, allowing to provide \emph{service-level} %
transparency information in line with the requirements from the GDPR identified in sec.~\ref{sec:legal_req}. \emph{System-wide} transparency information, in turn, are subsequently addressed in sec.~\ref{sec:thub}. We publish our contributions along with extensive documentation as open source software.\footnote{See \url{https://github.com/PrivacyEngineering/tira}}

\subsection{Personal Data in OpenAPI specifications}
\label{sec:openapi_structure}

An OpenAPI specification describes a RESTful API as a whole, including many details unrelated to the processing of personal data. Hence, the first step towards a transparency-specific OpenAPI extension consists of identifying where and how personal data becomes relevant and where respective transparency information is thus to be incorporated in an OpenAPI specification document. %
We refer to those parts of the specification that describe consumed or exposed personal data as \emph{PD indicators} herein.\footnote{For the sake of understanding and clarity: The OpenAPI specifications may under no circumstances contain personal data records (\enquote{Jane Doe}), but only \emph{PD indicators} (\enquote{Name}).}

The most relevant part with regards to \emph{PD indicators} is the \texttt{paths} section. 
All routes and requests provided by the described service are listed here. 
In the paths object, instances of \texttt{pathItem} can be defined, describing a single URL a service accepts requests for. 
Such a \texttt{pathItem} can consist of several \texttt{operations} (GET, POST, etc.) that map to HTTP request methods.
For each operation, in turn, all entities that make up a request and its response can be described.
These include request bodies, responses, headers, cookies and parameters, whereas OpenAPI consolidates path parameters (e.g. \lstinline|domain.tld/{user_id}|), query parameters (e.g \lstinline{/path?user_id=123}), headers and cookies under the \texttt{Parameter} class. All these may possibly specify the consumption or exposure of personal data and are therefore candidates for \textit{PD indicators}.\footnote{For more details see \url{https://github.com/PrivacyEngineering/tira/blob/main/docs/PD_INDICATORS.md}}

\subsection{Extending OpenAPI}\label{ssec:extending-openapi}
Having identified the locations where \emph{PD indicators} may reside, we introduce a custom extension for OpenAPI documents. Implementing such an extension is a prerequisite for actually expressing transparency information via a vocabulary. This vocabulary will be embedded into the extension and will be covered in detail in sec.~\ref{sec:vocabulary}. Our extension enables developers to:

\begin{itemize}
    \item[\circled{1}] Declare data fields processed by a RESTful service as personal data indicator as such
    \item[\circled{2}] Further annotate personal data indicators with transparency information regarding the processing done by the documented service
    \item[\circled{3}] Annotate properties of the service itself that are relevant with regards to transparency obligations
\end{itemize}

OpenAPI extensions can be of all primitive types, \texttt{null}, an \texttt{array} or an \texttt{object}, allowing for versatile custom extensions. We will use %
them to enrich the existing documents with transparency-related information. Moreover, in order to allow several extension types and for preventing overlaps with other arbitrary extensions, we define a namespace called \texttt{x-tira}. %
Hereafter, we describe how we leverage the extension to fulfill the tasks mentioned above.

To \circled{1} \textbf{declare any data field as personal data indicator} using \texttt{x-tira}, it is sufficient to include an \texttt{x-tira} extension instance of any form inside the corresponding \texttt{schema}. 
The recommended way is to set the custom boolean \texttt{x-tira} to \texttt{true} inside a \texttt{schema}. %
If a whole \texttt{schema} is marked this way, no specific \textit{PD indicators} inside the given \texttt{schema} would need to be declared. %
If transparency information is declared at higher hierarchical levels of the document, a subordinate \texttt{schema} will automatically inherit the given property or can possibly override it. Listing~\ref{lst:personal-data-weight-x-tira-only} exemplifies the declaration of \textit{PD indicators} in the context of our application scenario, describing the API endpoint the fitness devices %
communicate with. 

Alternatively, an \texttt{x-tira} object may also be declared at the root level of a document. Then, declared transparency information refers to the whole API endpoint. This may be useful when, for example, the hosting situation of the service itself has an influence on transparency obligations -- e.g. when the service is hosted outside of the EU jurisdiction or is provided by a third party (see the two gray-colored service providers in our application scenario).
Complementary attributes or properties of a \texttt{schema} for which the declaration shall not apply can be exempted by invoking \texttt{x-tira-ignore} (see listing~\ref{lst:personal-data-weight-x-tira-only}).

\begin{lstlisting}[
    language=yaml,
    basicstyle=\tiny,
    frame=single,
    float=tp,
    caption={
        [A schema indicating personal data being processed.]
        {
        An OpenAPI \texttt{schema} definition of the object \texttt{Weight} which is marked as \textit{PD indicator}.}
        },
    captionpos=b,
    label={lst:personal-data-weight-x-tira-only}
]
components:
  schemas:
    Weight:
      x-tira: true             # Declared as PD indicator
      type: "object"
      required: 
        - weight
        - day
      properties: 
        weight: 
          type: "number"
          format: "float"
        submission: 
          type: "string"
          format: "dateTime"
        log-level: 
          type: "string"
          x-tira-ignore: true   # Excluded from marking (not personal data)
\end{lstlisting}

\begin{lstlisting}[
    language=yaml,
    basicstyle=\tiny,
    frame=single,
    float=t,
    caption={
        Example of the \texttt{RetentionTime} vocabulary element:  A
        time-span is defined by setting values for years, months and days. 
        When storage is volatile or has no explicit limit, this is also configurable.
        Whether a review is happening -- as imposed by the GDPR -- and in which frequency it is performed can also be defined. 
        This example shows a storage period of ten years whose compliance is controlled daily. 
    },
    captionpos=b,
    label={lst:vocab:retention_time}]
x-tira: 
  retention_time:
    days: null
    months: null
    years: 10
    # volatile: true
    # no_limit: true
    periodic_review: true
    review_frequency: 
      days: 1
      # months: null
      # years: null
\end{lstlisting}

\subsection{Transparency vocabulary} \label{sec:vocabulary}

Having established how \textit{PD indicators} can be declared as such in an OpenAPI specification, we will now continue with how \circled{2} \textbf{further transparency-relevant information can be  expressed} in line with requirements from to the GDPR. For this purpose, we introduce a transparency vocabulary that is integrated into OpenAPI documents via the presented \texttt{x-tira} extension.
Doing so allows developers to annotate %
data processing activities of a given microservice. By design, the vocabulary closely relates to the legal requirements from the GDPR, which were laid out in sec.~\ref{sec:legal_req} (see Req.~1). The vocabulary represents all information that realistically can be expressed by developers (see Reqs.~2 and 5), hence representing their perspective. %

If a schema (see listing~\ref{lst:personal-data-weight-x-tira-only}) is marked as \textit{PD indicator}, our extension will instantiate a \texttt{Tira::PersonalDatum} object related to it. Attached to this object can be instances of \texttt{Tira::TransparencyProperty} which have a type. Each of these, in turn, represents one vocabulary entity, which is represented by a subclass of the \texttt{Tira::TransparencyProperties} class. In general, \texttt{TransparencyProperties} can be attached to a \textit{PD indicator} in several ways within an OpenAPI document: Either they are incorporated directly by extending the schema or they might as well be integrated on a higher level, e.g. for operations or pathItems, in which personal data is obtained or passed on. Alternatively, at document level, \circled{3} \textbf{transparency properties of a whole service are annotated}, e.g. for external services (see Reqs. 2 and 3). %

To illustrate how the vocabulary is used in practice, we recall our application scenario (see sec.~\ref{ssec:running-example}). We might want to further annotate the API specification of our PostgreSQL database to show exemplary behaviour. In listing~\ref{lst:vocab:retention_time} we show how a \textit{storage period} can be expressed fully compliant with table~\ref{tab:information-categories}. First of all, the storage period itself can be set. Developers can then express for how long the documented endpoint saves data via \texttt{days}, \texttt{months}, and \texttt{years} fields. If the storage is volatile or unlimited in time, this can also be set in respective fields. Moreover, the GDPR also allows storage time limits to be ensured by periodic reviews (see Recital 39). Whether these reviews are taking place and, if that is the case, in which frequency %
can also be set in the storage period vocabulary element. Taken as whole, the extended API specification effectively communicates the required transparency information with regards to storage periods.

For now, we took the vocabulary element of \texttt{RetentionTime} as an extensive example.
Beyond this, we also introduce vocabulary elements for \texttt{Recipients}, \texttt{Third Country Transfers}, \texttt{Special Category} (Art. 9 GDPR), \texttt{Profiling}, and \texttt{Purposes} (here we integrated YaPPL \cite{cite_yappl}).
All of these vocabulary elements consist of curated fields for declaring \textit{all} relevant information according to table~\ref{tab:information-categories} (see Req.~1) along with \textit{PD indicators}. Eventually, all vocabulary items are managed by the \textit{TransparencyHub} for further aggregation. Due to space constraints, %
a more extensive explanation of each item is given in our documentation.\footnote{See \url{https://github.com/PrivacyEngineering/tira/blob/main/docs/VOCABULARY.md}}

\section{Managing system-wide transparency information with \textit{TransparencyHub}}
\label{sec:thub}
As laid out above, the transparency information provided via our OpenAPI extension on a per-service level also need to be integrated into a single, overall view comprising all relevant services and their interdependencies. 
In addition, they also need to be enriched with system-wide information (see table \ref{tab:information-categories}) for being able to fulfill all transparency requirements of the GDPR. These functionalities are provided by the \textit{TransparencyHub} we propose alongside our OpenAPI extension.

The \textit{TransparencyHub} implements an endpoint for a CI  workflow, that accepts and processes transparency-enhanced RESTful service descriptions (integrative functionalities),
analyzes and aggregates these (analytical functionalities), and
offers a transparency administration interface

(management functionalities). 
The \textit{TransparencyHub} was built using the popular \textit{Ruby on Rails} web-application framework, with the OpenAPI extension and vocabulary being implemented as separate modules each. %
We will now elaborate on the workflows provided by the \textit{TransparencyHub}.

\subsection{Aggregating transparency information}
\label{subs:thub:displaying_transparency_properties}

In essence, the \textit{TransparencyHub} collects and processes the transparency information provided via services' OpenAPI documents and translates these into consolidated representations and views, accessible via multiple dashboards: 
First, an index view for all registered internal and external services is created. 
One can, at a glance, check whether an OpenAPI specification is present and valid,
access an interactive \textit{SwaggerUI} representation\footnote{See \url{https://swagger.io/tools/swagger-ui/}} -- as \textit{Swagger} utilizes the OpenAPI format --
and view a history of specification changes and compare older versions (commit log). 
Services that do not process personal data can be viewed but reside in a separate index view.

\begin{figure}[t]
		\includegraphics[width=1\columnwidth]{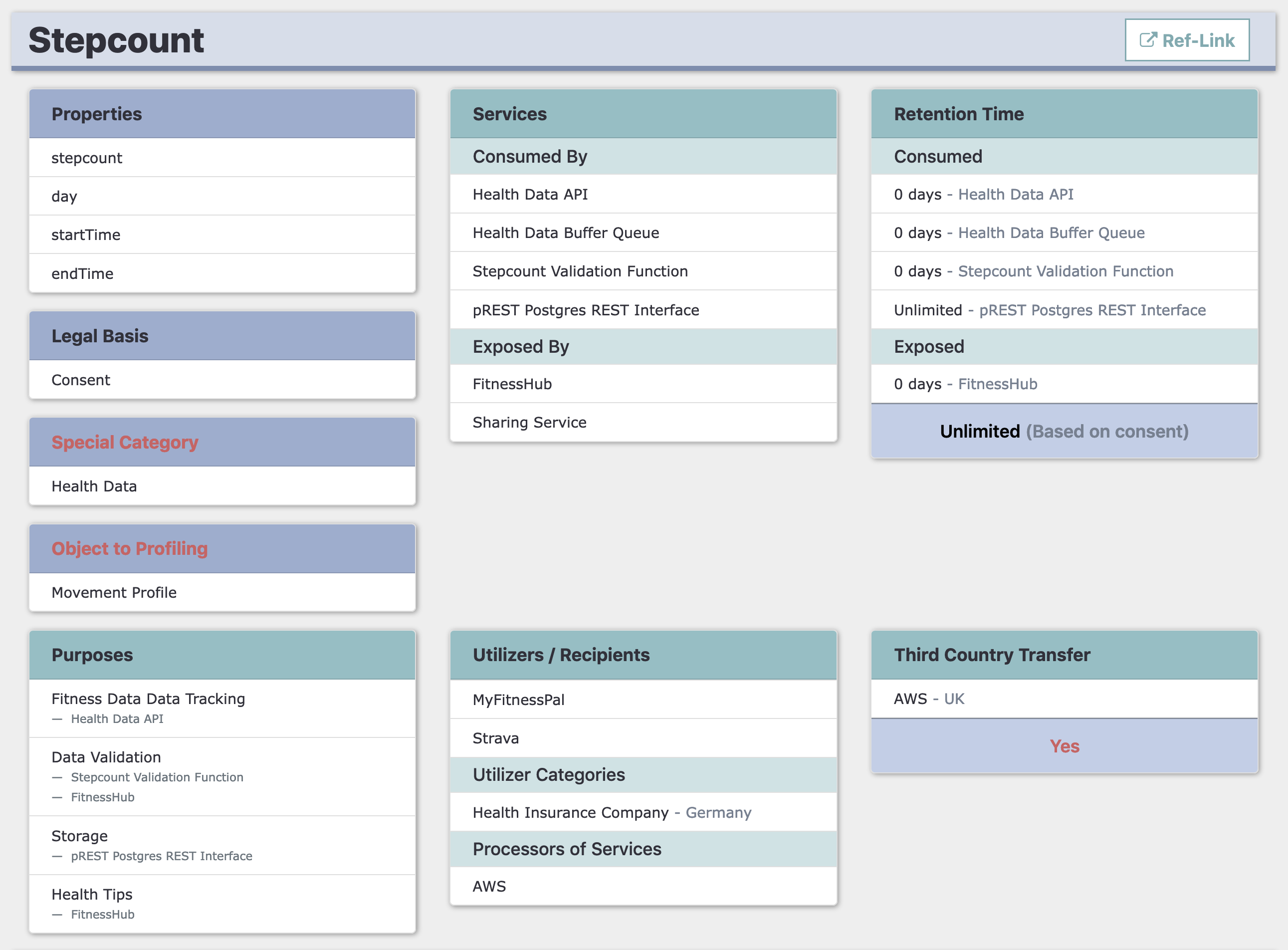}
		\caption[\textit{TransparencyHub} -- Aggregated Transparency Information for one Personal Datum]
		{
		    One of the generated dashboards: Aggregated transparency information for a \textit{PD indicator}
		    (here: \enquote{Stepcount} from a health data device). 
		  It provides an overview of services the data are processed by and all properties incorporated via the transparency vocabulary in OpenAPI documents are shown. 
		  Via the `Ref-Link', the respective reference in an OpenAPI document can be copied. All named services relate to our application scenario (see figure~\ref{fig:example-architecture}).
 		}
		\label{fig:thub-stepcount}
		\vspace{-0.3cm}
\end{figure}

Secondly, we provide a schema index including all entities that were marked as \textit{PD indicators} (see sec.~\ref{ssec:extending-openapi}). 
Such an aggregated view (see figure~\ref{fig:thub-stepcount}) accurately displays the gathered transparency information that were expressed via the vocabulary about \textit{PD indicators} and entire services.
To actually aggregate the distributed transparency information, \textit{TransparencyHub} uses aggregation functions to combine vocabulary elements of the same type regarding each \textit{PD indicator}.\footnote{E.g., the storage periods seen in figure~\ref{fig:thub-stepcount} were aggregated to an \enquote{unlimited} value based on the arithmetic maximum used as suitable aggregation function.}

In the same way, a comparable dashboard is present for all purposes that are present throughout all OpenAPI specifications. All purpose/service and purpose/\textit{PD indicator} dependencies become visible in there.
Analogously, a dashboard is generated for all data recipients or utilizers and respective categories thereof.

\subsection{Managing transparency information}\label{subs:thub:express_transparency_info}

From both a systems engineering and a legal perspective
not all transparency-related information that need to be expressed may directly relate to RESTful APIs. In addition, it would not be viable to have developers alone being responsible for specifying \emph{all} relevant transparency information. %
The \textit{TransparencyHub} hence does not only aggregate gathered transparency information, but also acts as transparency administration interface for the data controller (or data protection officer) and the entire system.

To begin with, the \textit{TransparencyHub} allows to keep track of OpenAPI documents of arbitrary services.
Therefore, services -- internal or external ones -- can either be registered manually (by uploading the description) or be added automatically via our CI/CD pipeline integration (see sec.~\ref{sec:ci-cd}). 
After services have been registered, their interconnections can be edited, defining all \textit{n:m} relations with regard to sending and receiving personal data. 
This will reveal the data flows within the system on a high level -- which is also displayed as a graph structure -- supporting a possible processing inventory. 
If third parties would also make use of extended OpenAPI documentations, %
these can be integrated in the analysis just as any internal service -- allowing to achieve \enquote{transitive transparency} across multiple %
companies and business domains.

Additionally, the \textit{TransparencyHub} serves as administration point for system-wide transparency information such as contact information, data protection officers, etc., which are required for fulfilling legal %
obligations (see table~\ref{tab:information-categories}). %
Also the legal bases, for which we provide templates, and Art. 9 processing scenarios can be edited and described further. 
Besides, all utilizers and purposes can each be grouped if they concern the same processing activity. 
For a future version, we also consider directed acyclic graph structures for linking and building hierarchies between these entities and categories thereof such as described in related work \cite{cite_yappl}.

\subsection{Continuous Integration and Delivery of transparency information}
\label{sec:ci-cd}

To seamlessly integrate the practical application of our approach into modern, agile development practices and the respective tooling  (Req.~4), we provide a pipeline component that automatically %
processes transparency-enhanced service specifications %
with as little manual developer effort as possible.
For this purpose, we provide a CI component that informs the \textit{TransparencyHub} about changes to services processing personal data.

We chose an integration into Git-based version control systems\footnote{We chose \textit{GitLab} for our implementation -- analogous functionalities are present for \textit{Github}, \textit{Bitbucket}, and \textit{Gitea}, among others.}. Therefore, we implemented an endpoint accepting and processing Git \texttt{push} events, which can be configured via Git webhooks present in all service repositories.
If a repository is yet unknown, the \textit{TransparencyHub} will automatically create a service representation, otherwise it continuously registers whether existing OpenAPI documents have been added (e.g., new \textit{PD indicators}) or changed (e.g., updated purpose definitions).%

Through this functionality, it is possible to configure all services of an organization to use the introduced technologies and enable a transparency-aware system via just one initial setting.

\section{Discussion \& Conclusion}
\label{sec:conclusion}

In this paper, we presented TIRA, a toolbox for achieving GDPR-related transparency in RESTful application architectures developed and operated following agile, DevOps-oriented practices and paradigms. Our contribution comprises, first, an OpenAPI extension and respective vocabularies specifically tailored to the GDPR's transparency requirements. Second, TIRA also includes a set of higher-order tools for collecting and aggregating respective transparency information from the descriptions of multiple, interdependent services %
and for integrating our approach into established CI/CD-pipelines. %
Tightly interwoven with each other, these components allow to harness transparency information in a bottom-up, developer- and service-focused manner. The so-generated transparency information pave the way for novel approaches for fulfilling regulatory transparency requirements in line with the widely established givens 
of modern, real world information systems engineering. 

Our approach significantly differs from other proposals made so far in various respects. First and foremost, TIRA is based on an exhaustive analysis of legal transparency requirements and explicitly tailored to these from ground up. Insofar, it significantly distinguishes from other approaches %
(e.g., \cite{cranor2002web, prime}) which often lack the explicit alignment to legally mandatory transparency requirements \cite{gerl2018critical, dark}. Second, we consciously follow an annotation-based approach instead of employing, for instance, information flow control \cite{pasquier2016information} or static code analysis \cite{hjerppe2019annotation} for determining what personal data a service consumes or exposes, how long it is stored, etc. Here, we consciously take an \enquote{optimistic} stance on developers and data controllers -- which we aim to support in fulfilling their duties instead of trying to prevent misbehavior -- and explicitly value practical applicability over tamper- or concealment-proofness. In matters of integrating the procedural and organizational givens of agile, DevOps-based information systems engineering from the outset and rigidly translating them into corresponding technical mechanisms, in turn, our approach is -- %
to the best of our knowledge -- quite unique in the domain of privacy engineering so far. We do, however, foresee more such activities dedicatedly integrating organizational givens %
into the design of technical privacy mechanisms to gain traction in the future.

Of course, our approach is currently still subject to some limitations. In particular, TIRA is so far confined to RESTful APIs with respective OpenAPI specifications. Even though components without such specifications could be encapsulated with additional wrappers
, doing so will not be viable in many real-world enterprise systems. Alternative service specification mechanisms -- such as the RESTful API Modeling Language (RAML), 
API Blueprint, 
GraphQL schemas, 
or Docker Compose
service configurations -- should thus also be integrated into TIRA in the future. Another line of desirable future extensions regards capabilities for reflecting and expressing advanced concepts such as, for instance, hierarchical vocabularies of data- or purpose-categories \cite{pandit2019creating, bonatti2018data}. Finally, integrating TIRA with recently proposed formats for representing and communicating transparency information across organizational boundaries  \cite{gruenewald2021, Gerl2018} is also an obvious target for future activities.

Apart from these current limitations, however, TIRA provides a paradigmatically novel approach for fulfilling transparency obligations in line with privacy regulations as well as the realities and givens of modern information systems engineering in practice. It is explicitly designed to lower data controllers' efforts and heightens their capabilities to fulfil their transparency-related duties.

Last but not least, TIRA thereby consciously works towards data controllers' obligation to actually apply technical transparency measures like the ones presented herein under legal provisions for privacy / data protection by design, such as Art. 25 of the GDPR. Beyond introducing a novel technical approach in and by itself, TIRA thus also demonstrates the wider prospects of applied privacy engineering beyond anonymization and security.

\section*{Acknowledgment}
     The work behind this paper was partially conducted within the project DaSKITA, supported under grant no. 28V2307A19 by funds of the Federal  Ministry of Justice and Consumer Protection (BMJV) based on a decision of the Parliament of the Federal Republic of Germany via the Federal Office for Agriculture and Food (BLE) under the innovation support program.

\bibliographystyle{IEEEtran}
\bibliography{references}

\end{document}